# Epitaxial fabrication of AgTe monolayer on Ag(111) and the sequential growth of Te film

Haoyu Dong [1], Le Lei[1], Shuya Xing [1], Jianfeng Guo[1], Feiyue Cao[1], Shangzhi Gu[1], Yanyan Geng[1], Shuo Mi[1], Hanxiang Wu[1], Yan Jun Li,[2] Yasuhiro Sugawara,[2] Fei Pang[1], Wei Ji[1], Rui Xu[1] and Zhihai Cheng[1*]

[1]*Beijing Key Laboratory of Optoelectronic Functional Materials & Micro-nano Devices, Department of Physics, Renmin University of China, Beijing 100872, People's Republic of China*

[2]*Department of Applied Physics, Graduate School of Engineering, Osaka University, 2-1 Yamadaoka, Suita, Osaka 565-0871, Japan*

**Abstract:** Transition-metal chalcogenides (TMCs) materials have attracted increasing interest both for fundamental research and industrial applications. Among all these materials, two-dimensional (2D) compounds with honeycomb-like structure possess exotic electronic structures. Here, we report a systematic study of TMC monolayer AgTe fabricated by direct depositing Te on the surface of Ag(111) and annealing. Few intrinsic defects are observed and studied by scanning tunneling microscopy, indicating that there are two kinds of AgTe domains and they can form gliding twin-boundary. Then, the monolayer AgTe can serve as the template for the following growth of Te film. Meanwhile, some Te atoms are observed in the form of chains on the top of the bottom Te film. Our findings in this work might provide insightful guide for the epitaxial growth of 2D materials for study of novel physical properties and for future quantum devices.

* To whom correspondence should be addressed: feipang@ruc.edu.cn    zhihaicheng@ruc.edu.cn

**Introduction**

The intriguing properties of two-dimensional (2D) materials have attracted great attention and become a hot and expanding topic since the discovery of graphene[1-4]. A lot of materials including the graphene[5-9] family (e.g., silicene, germanene, antimonene, stanene and borophene)[10-14] and other materials with honeycomb structure have been extensively studied both for fundamental research and industrial applications in the very recent years. It has been proposed that some 2D compounds with honeycomb structure have exotic electronic characters[15-20]. Also, in some honeycomb structure materials, electrons hopping can realize quantum anomalous Hall effect[21-23], which may lead to industrial applications.

As a large family of 2D compound materials, it is particularly attractive to explore the layered transition-metal chalcogenides (TMCs)[24-25]. Unlike most sandwich structure TMCs, some TMCs with planar honeycomb structures such as CuSe and AgTe may lead to topologically nontrivial electronic states due to their symmetry, including crystal symmetry and time-reversal symmetry[26-27]. Monolayer CuSe and AgTe in planar honeycomb structure yields comprehensive evidence to possess Dirac node-line fermions (DNLFs) protected by mirror reflection symmetry[28-29]. It is also confirmed experimental and theoretical that honeycomb monolayer AgTe for an orbital angular momentum-based origin of the Rashba effect[30]. For TMCs materials, the intrinsic defects not only play an important effect on the structure and property, but also could host interesting electronic states[31]. Therefore, the investigation of defects can be of great scientific and application relevance.

Here, we successfully synthesized high-quality flat TMC AgTe monolayer in honeycomb structure by direct depositing and annealing Te on Ag(111) surface, which is further investigated by reflection high-energy electron diffraction (RHEED) and scanning tunneling microscopy (STM). Two types of domains and three kinds of intrinsic defects are observed and studied, in which the gliding twin-boundary is caused by two different domains on each side. Two types of Te are found when we continue to deposit Te on AgTe monolayer, the bottom-Te form a rough ordered hexagonal phase, while the top Te form various short chains on the top of bottom Te film.

**Methods**

The Ag (111) (MaTeck, Germany) substrate was cleaned by cycles of Ar$^+$ sputtering and annealing at 700K, until a clean and uniform surface was obtained. The experiment was performed on Ag (111) substrate in a molecular beam epitaxy chamber operating at ultrahigh vacuum (UHV, base pressure $2 \times 10^{-10}$ torr). In order to get a sample with a small Te coverage, high-purity Te was evaporated from a standard Knudsen cell at 520K for a few seconds, with a substrate temperature of 320K. To fabricate flat AgTe, high-purity Te was evaporated at 520K for 2 minutes, then annealing at 770K for 1 hour. The growth process was monitored by reflection high-energy electron diffraction (RHEED). After each growth, the films were transferred through a high vacuum transfer chamber into the analysis chamber for STM measurements.

All the STM measurements were conducted in an ultra-high vacuum system equipped with a combined low-temperature STM at a base pressure of $2 \times 10^{-10}$ torr. All experiments were performed at 9K and with chemical etched W tip. The tips were calibrated on a clean Au (111) surface prepared by repeated cycles of sputtering with argon ions. The single crystals of Au (111) (MaTeck, Germany) were cleaned by repeated cycles of argon ion sputtering and annealing at 850 K.

## Results and Discussion

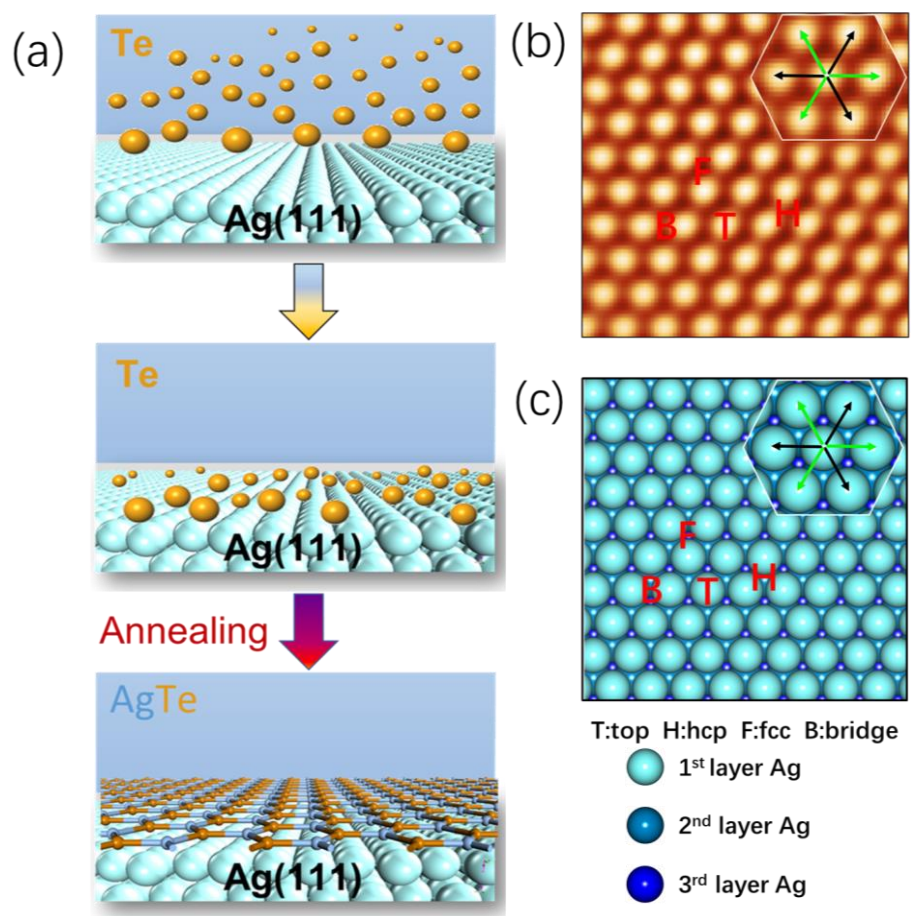

**Fig. 1. Schematic of AgTe monolayer preparation on Ag(111) surface.** (a) Schematic of Te atoms deposited on Ag(111) surface and the epitaxial fabrication of AgTe by annealing. (b,c) STM image (b) and atomic model (c) of Ag(111) surface. The high-symmetric sites are highlighted by F (fcc), H (hcp), B (bridge) and T (top) in both (b) and (c). Cyan, steelblue and darkblue balls represent Ag on the first layer, second layer and third layer, respectively. Inset: three-fold symmetry of Ag(111) surface.

Figure. 1(a) schematically shows the fabrication process of monolayer AgTe film on Ag(111) substrate, mainly including the epitaxial growth of Tellurium atoms on Ag(111) substrate and the following annealing process. The Ag(111) substrate was first cleaned by several cycles of sputtering and annealing process. The clean surface terraces were obtained and atomically resolved in the STM image of Fig. 1(b). Both the surface Ag atoms and FCC/HCP hollows were clearly imaged. For the following adsorption of Tellurium atoms, four possible high-symmetric positions were marked in the structural mode of Ag(111) surface, as shown in Fig. 1(c). For the Ag atoms of top layer, the inversion-symmetry is preserved at the Top sites but not at the Hollow sites.

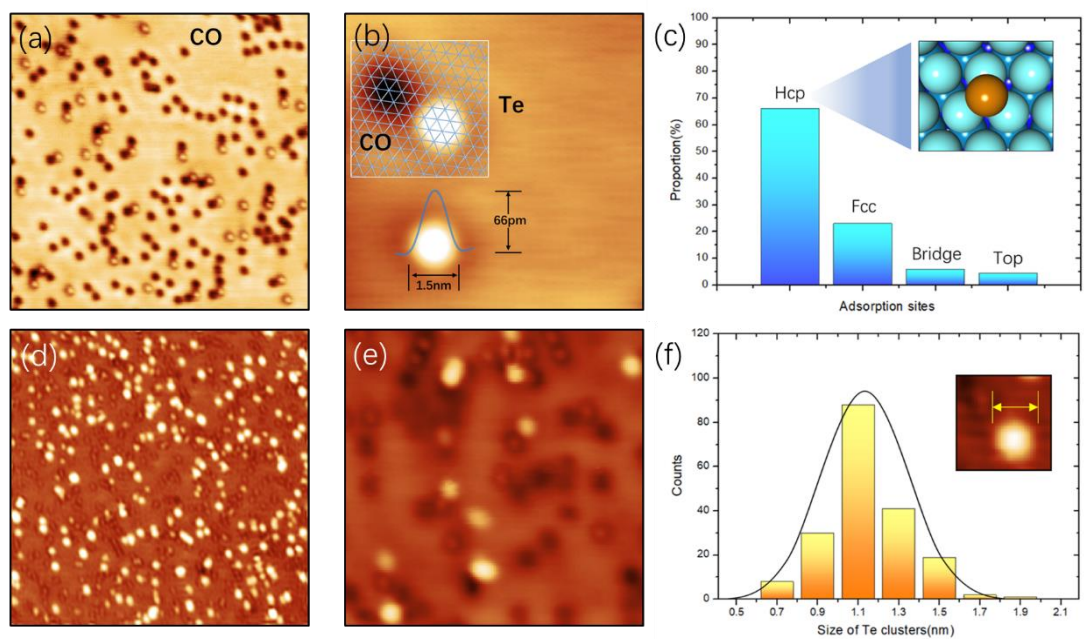

**Fig. 2. STM measurements of Te adatoms and clusters on Ag(111) surface**. (a) Large-scale (20nm× 20nm, 0.8V, 100pA) and (b) high-resolution (6.5nm×6.5nm, 0.3V, 100pA) STM images of Te adatoms on Ag(111) surface. The adsorption sites of Te adatoms were determined with the surrounding reference CO molecules. (c) Histogram of the identified adsorption sites of Te adatoms. (d) Large-scale (40nm×40nm, -1.5V, -100pA) and (e) high-resolution (10nm×10nm, 0.4V, 100pA) STM images of Te clusters on Ag(111) surface. (f) Histogram of the measured sizes of Te clusters.

After a very small amount of Te atoms is deposited on the clean Ag(111) substrate kept at room temperature, the Te adatoms randomly adsorbed on the Ag(111) surface in the monomer way, as shown in the STM image of Fig. 2(a). The carbon monoxide (CO) molecules were purposely introduced on the Ag(111) surface to determine the adsorption sites of Te adatoms[32-35], in which the CO molecules appear as the round dark depressions. The adsorption sites of Te adatoms were determined with the surrounding reference CO molecules adsorbed at the top sites of Ag(111) surface, as shown in Fig. 2(b). The statistical results of the determined adsorption sites of Te adatoms are shown in the histogram of Fig. 1(c). In general, the Te adatoms prefer to be adsorbed on the Hollow sites, especially the Hcp hollow sites. The apparent size and height of Te adatom is roughly estimated at 1.5 nm and 66pm, respectively. As the coverage of Te adatoms increases, both the Te clusters and adatoms were observed in the large-scale STM image of Fig. 2(d). The apparent size of Te clusters were statically measured and shown in the histogram of Fig. 2(f). It is noted that the pristine surface states of pristine Ag(111) substrate were clearly modified by the Te adatoms and clusters, indicating their strong bonding with the surface atoms of Ag(111).

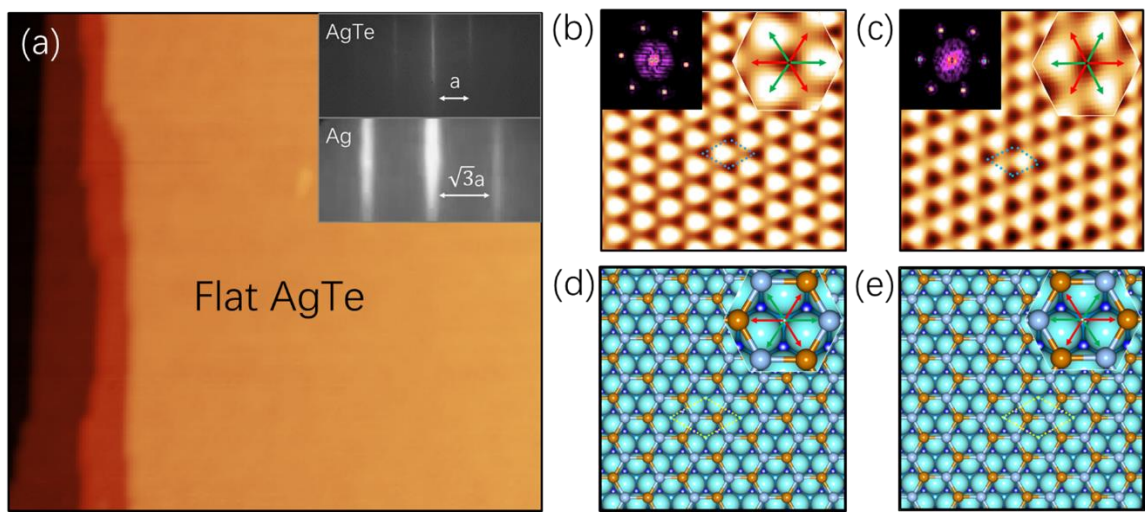

**Fig. 3. STM measurements of AgTe monolayer on the Ag(111) substrate.** (a) Large-scale STM image (50nm × 48nm, -2V, -100pA) of flat AgTe monolayer. Inset: RHEED of AgTe monolayer and Ag substrate. (b,c) Atomic-resolution (4nm×4nm, -1.6V, -60pA) STM images of AgTe monolayer. (d,e) The corresponding atomic models of AgTe monolayer in (b,c). The insets of FFT and symmetry analysis confirm the inversion-symmetric equivalent structures of AgTe in (b) and (c).

After the deposition of Te atoms on the Ag(111) surface, the following annealing process was performed to form the AgTe monolayer film[36-37]. A large-scale STM image of the formed AgTe monolayer is shown in Fig. 2(a). The inset RHEED patterns indicate the ordered (√3×√3) superstructure of AgTe monolayer with the underlying Ag(111) substrate. Two kinds of atom-resolved STM images of AgTe monolayer were observed in the STM measurements, as shown in Fig. 3(b) and 3(c). The atoms are arranged in a periodic superstructure with three-fold symmetry. Their lattice constant is 5.0 Å, which is √3 times the underlying Ag(111) lattice constant (2.88 Å) in agreement with the RHEED results. There are two types of atoms with different apparent heights that were observed in the STM images. The higher (lower) atoms appear as bright (dark) triangles and are determined as the Ag (Te) atoms of the AgTe monolayer, as shown in the corresponding atomic model of Fig. 3(d) and 3(e). It is noted that the AgTe monolayers of Fig. 3(b) and 3(c) are inversion-symmetric equivalent on the Ag(111) surface, which could form the boundaries when they meet each other during the growth process.

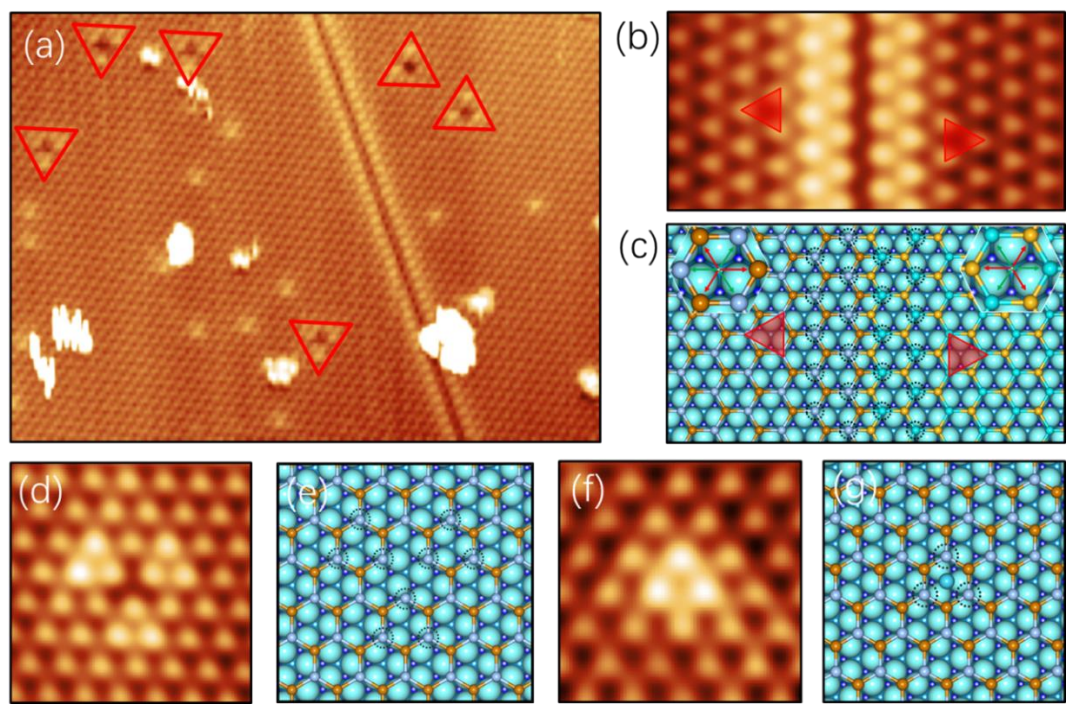

**Fig. 4. Intrinsic defects of AgTe monolayer.** (a) Large-scale STM image (25nm×18nm, -1.4V, -100pA) of AgTe defects. (b,d,f) Atomic-resolution STM images of (b) AgTe line defect (5nm×2.5nm, -1.6V, -60pA), (d) point defect (3.5nm×3.5nm, -1.6V, -60pA) and (f) interstitial atom (2.5nm×2.5nm, -1.6V, -60pA) in AgTe film. (c,e,g) The corresponding atomic models of (b,d,f).

The imperfect structures, such as boundaries and atomic defects, play an important effect in modification of their inherent properties for the monolayer 2D material. The intrinsic defects of AgTe monolayer were further investigated, as shown in Fig. 4. Three kinds of defects were observed in the large-scale STM image of Fig. 4(a). The one-dimensional (1D) boundary between the two inversion-symmetric AgTe domains is shown in the STM image of Fig. 4(b) and the corresponding atomic model of Fig. 4(c). The apparent heights of the Ag atoms in the four silver atom rows (highlighted by the dotted circles) are clearly larger than those of normal Ag atoms in the defect-free regions. It is also noted that this kind of domain boundary has a gliding plane, named gliding twin-boundary, which is determined by the underlying Ag(111) substrate and different from the mirror twin-domain boundaries in the MBE-grown TMD monolayers. The mirror twin-boundaries have been reported in various transition-metal dichalcogenides (TMD) monolayer, which is metallic and could host exotic Tomonaga-Luttinger liquid state[38]. The formation of gliding twin-boundary in AgTe monolayer is due to the interplay between the inversion-symmetric domains of AgTe monolayer and the uniform underlying Ag(111) substrate. Recently, it has been theoretically reported that the free-standing monolayer AgTe hosts two Dirac node lines[29]. When further considering the SOC effect, the possible topologically nontrivial properties were proposed due to the gap opening at Dirac points. This gliding twin-boundary could host interesting electronic states. The remaining two kinds of point atomic defects were shown in the STM images of Fig. 4(d) and 4(f). The atomic defect in Fig. 4(d) appears as a single Te vacancy and the nine surrounding Ag atoms with larger heights, which are marked in the atomic model of Fig. 4(e). The atomic defect in Fig. 4(f) appears as three higher Ag atoms surrounding the hollow center, which is proposed as the intercalated silver adatom between the AgTe monolayer and the underlying Ag(111) substrate.

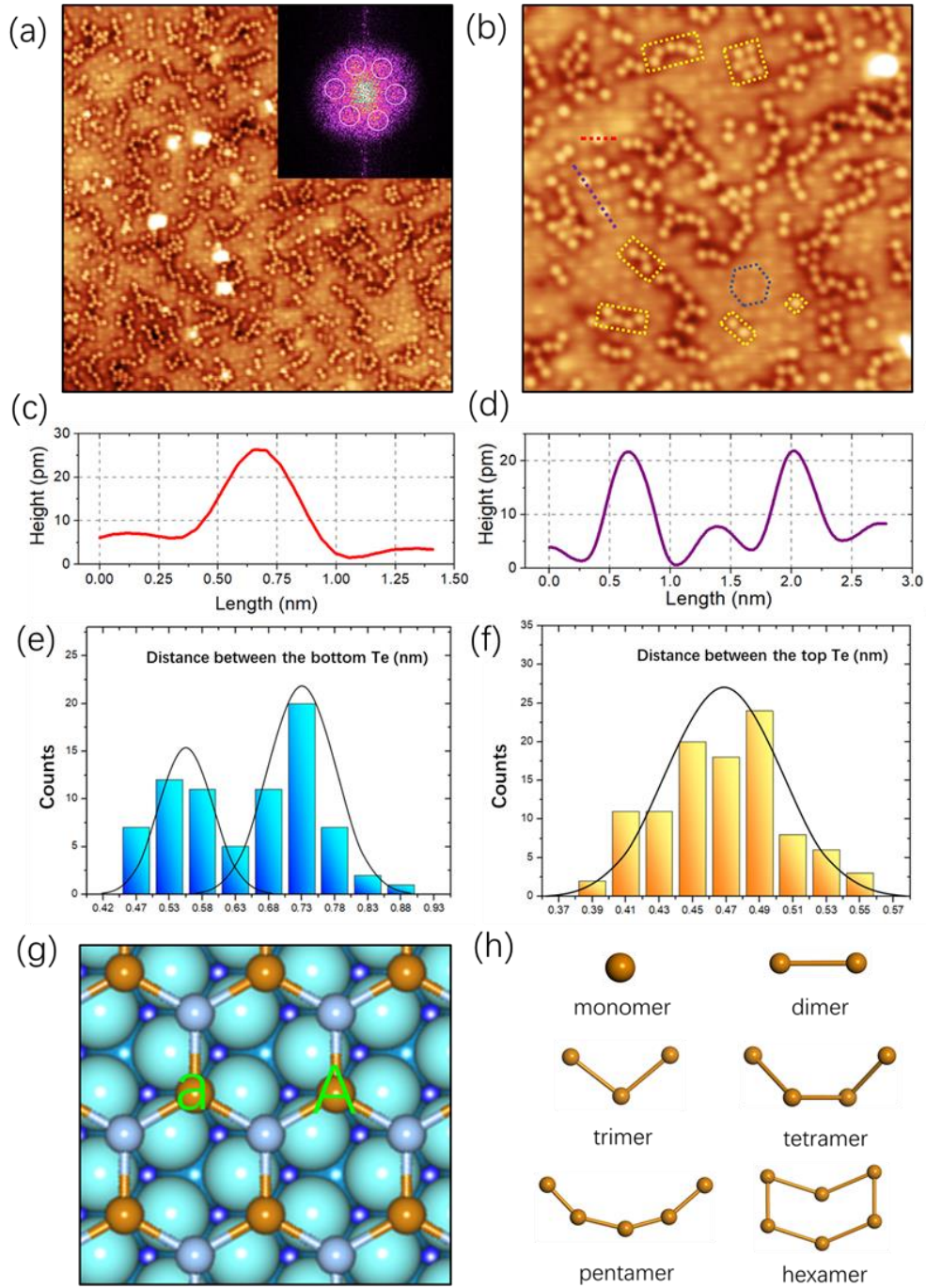

**Fig. 5. Self-assembled structures of Te on the AgTe monolayer.** (a) Large-scale (30nm×30nm, -0.8V, -100pA) and (b) zoomed-in STM images (15nm×15nm, -0.8V, -100pA) of two kinds of Te on the surface. Inset: FFT of STM image. (c,d) Line profiles of red (c) and purple (d) dashed lines in (b). (e,f) Histogram of the apparent inter-atom distance within the bottom Te (e) and top Te (f). (g) Atomic model of monolayer AgTe on Ag(111) surface. "a" and "A" shows two adjacent Te sites in AgTe monolayer. (h) Schematic of Te chains from Te monomer to Te hexamer.

The element tellurium ordinarily forms a crystal structure consisting of atomic chains that spiral along one of the axes of the crystal lattice, which can be assumed as a bundle of 1D nanowires. Recently, tellurene, the 2D monolayer of Te, has been theoretically predicated that it can exist in the hexagonal structure of stable 1T-$MoS_2$-like and metastable 2H-$MoS_2$-like monolayer[39-42]. Here, the epitaxial AgTe monolayer could serve as the template for the following growth of the bottom Te film on it.

The large-scale and high-resolution STM images of the epitaxial Te film on the AgTe monolayer template are respectively shown in Fig. 5(a) and 5(b). Two kinds of Te atoms were confirmed according to their different apparent heights in the STM measurements, as shown in Fig.5(c) and 5(d). The bottom Te atoms form a rough ordered hexagonal phase, which is indicated by the inset FFT pattern of Fig. 5(a) and marked with blue dotted hexagon in Fig. 5(b). The apparent inter-atom distance within the bottom Te structures were statically measured and shown in the histogram of Fig. 5(e). Two specific inter-Te distances of ~5.3 Å and ~7.3Å were discovered, in which the short one is close but larger than the inter-Te distance of AgTe monolayer indicated by "a" and "A" sites in Fig. 5(g). According to their multivalence nature, the bottom Te atoms were proposed to bond with the constituent Te atoms of AgTe monolayer. It seems that there are repulsive interactions between them to make them in the tilted configuration separately from each other.

The top Te atoms form various short chains on the top of the bottom Te film. The typical short chains are marked by the yellow dotted rectangles in Fig. 5(b). The apparent inter-atom distance within the chains were statically measured and shown in the histogram of Fig. 5(f). The specific inter-Te distances of ~4.7 Å were discovered and in agreement with the inter-atom distance of helical atomic chains in the ordinary stable crystal structure. Figure 5(h) shows the corresponding three-dimensional (3D) structural models of these typical structures. Both relative straight and curved chains were observed, indicating that the angles of Te-Te covalent bonds are soft due to the multivalence nature of Te atoms.

**Summary**

In conclusion, we have systematically investigated the characters and defects of honeycomb structure monolayer AgTe and AgTe monolayer served as the template for the following growth of Te film on it. By direct tellurization of Ag(111), flat monolayer AgTe is synthesized and two domains are observed. Three kinds of intrinsic defects, including line defect as well as two types of point defects, are discovered and analyzed according to symmetry. By depositing Te atoms on AgTe substrate, hexagonal phase bottom Te and short chains top Te are formed and analyzed statistically. Further investigations are expected for monolayer AgTe and its intriguing properties as a promising system for fundamental research and future applications in electronic devices.

**Acknowledgments:** This project is supported by the Ministry of Science and Technology (MOST) of China (No. 2016YFA0200700),the National Natural Science Foundation of China (NSFC) (No. 61674045,61911540074), the Strategic Priority Research Program and Key Research Program of Frontier Sciences (Chinese Academy of Sciences, CAS) (No. XDB30000000, No. QYZDB-SSW-SYS031), Grant-in-Aid for Scientific Research from Japan Society for the Promotion of Science (JSPS) from the Ministry of Education, Culture, Sports, Science, and Technology of Japan (JP16H06327, JP16H06504, JP17H01061, JP17H010610), and Osaka University's International Joint Research Promotion Program (J171013014, J171013007, J181013006, Ja19990011). Z. H. C. was supported by the Fundamental Research Funds for the Central Universities and the Research Funds of Renmin University of China (No. 21XNLG27).